# Investigation of the Electronical and Optical Properties of Quantum Well Lasers with Slightly Doped Tunnel Junction

Yajie Li [1,2], Pengfei Wang [1,2], Fangyuan Meng [1,2], Hongyan Yu[1], Xuliang Zhou [1], Huolei Wang[1,3], Jiaoqing Pan[1,2,*]

[1]Key Laboratory of Semiconductor Materials Science, Institute of Semiconductors, Chinese Academy of Science, Beijing 100083, China.
[2]College of Materials Science and Opto-Electronic Technology, University of Chinese Academy of Sciences, Beijing 100049, China.
[3]Department of Applied Physics and Materials Science, California Institute of Technology, Pasadena, California 91125, USA
*Corresponding author: jqpan@semi.ac.cn

**We experimentally investigate and analyze the electrical and optical characteristics of GaAs-based conventional quantum well laser diodes and the quantum well laser diodes with slightly-doped tunnel junction. It was found that TJ LD show a nonlinear S-shape I-V characteristic.It was also found that the internal quantum efficiency measured by 21% and 87.3% for the TJ LD and the conventional LD, respectively. Furthermore, compared with the conventional LD, it was found that we could achieve 15 nm broadband spectrum from the TJ LD due to lasing dynamics reflects the current dynamics. The results may also lead to the realization of more applications.**

Semiconductor laser diodes (LDs) are used at present in a variety of fields in science and technology. Tunnel junction (TJ), which was proposed by Esaki [1], has been widely used in many semiconductor LDs for different applications. Overview the TJ semiconductor LDs, there are mainly three different kinds of structures: First, reverse-biased TJs comprised of heavily doped p-type and n-type layers are located between adjacent LDs provides good electric contact [2]. This kind of semiconductor LDs can enhance the quantum efficiency and output optical power of lasers. The principle of these devices is to inject one electron-hole pair into the active regions to produce N times electron-hole pairs through cascaded TJs, where N is the number of the active regions. The internal quantum efficiency of this cascaded lasers can be much larger than unity while for conventional LDs. Second, dual-wavelength lasers based on the bipolar cascade lasers (BCL) with a similar structure also have been demonstrated [3]. The dual-wavelength BCL is composed of two different active regions with different multi-quantum-wells (MQWs) and the active regions are connected by a very thin reverse-biased TJ. The TJ requires materials with high n-type and p-type doping levels, low dopant diffusion, and large and gaps to avoid absorption. Third, the forward-biased TJ with a high doped n-type and p-type layers are located on the top of a LD [4], the negative differential resistance (NDR) of the TJ have been used in optical bistable devices, which are the key components in future optical switching and computing systems. The principle of these devices is the tunnel diode section provides a differential negative resistance which induces an electrically bistable characteristic when connected in series with the laser section and an external load. Thus the coherent light output from the device displays a negative differential optical response and bistable effects in response to an applied voltage. Optical bistable devices have very large application prospects in the all-optical data switch [5], all-optical signal amplification [6], optical pulse width compression and signal shaping [7], all-optical signal reproduction [8] and many other signal processing. In practical applications, the optical bistable switch requires a high switching speed and a low forward voltage drop to increase the switching frequency and reduce switching losses and forward conduction losses. In the general case, the electrical bistability with two equilibrium states can be obtained in a thyristor heterostructure [9]. The thyristor structure has a variety of advantages for optical switch, such as low switching energy, low power consumption, fast response, high on/off contrast, and expansibility to two-dimensional monolithic array [10-13].

In these TJs, the high-doping of the layers composing TJ plays a crucial role for optimizing and engineering device properties. Recently the slightly-doped TJ LDs have been reported to achieve the broadband stimulated emission [14-15]. However, the mechanisms behind these principles are not revealed yet, furthermore, how the slightly-doped tunnel junction determine the laser performance also unrevealed in the reported literatures. In this letter, we report GaAs/InGaAs quantum well lasers with the slightly-doped p-n+ GaAs-GaAs reverse-biased TJ located below the upper cladding layer. The electronical and optical properties are investigated and analyzed compared with conventional LDs. Detailed fabrication procedures and the electro-optical properties of the fabricated LDs will also be discussed.

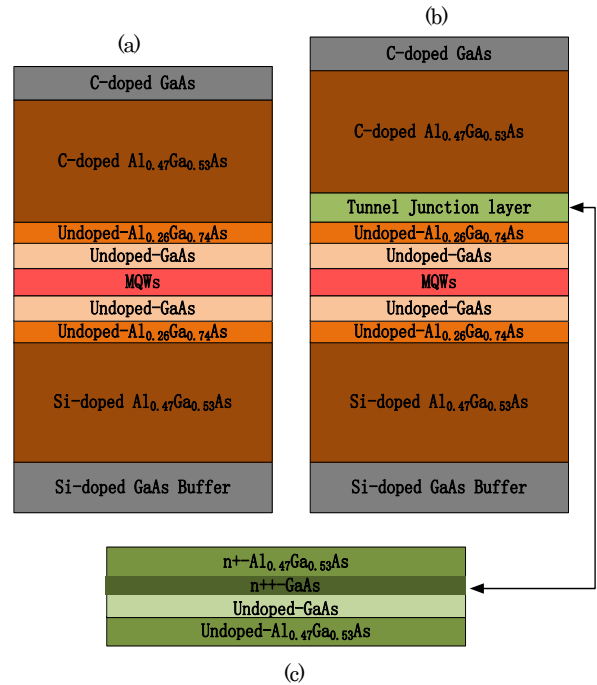

Fig. 1. Schematic diagrams of (a) the conventional LD, (b) the slightly-doped TJ LD, and (c) the slightly-doped TJ structure used in this study.

Samples used in this study were grown on a heavily n-doped GaAs substrate via Metal Organic Chemical Vapor Deposition (MOCVD) using only one step growth. Fig. 1. (a) and (b) schematically depict the structures of the conventional LD and the slightly-doped TJ LD, respectively. For the conventional LD, we first deposited a 400-nm-thick GaAs buffer layer, a 1.8-μm-thick Si-doped n-AlGaAs cladding layer and a 100-nm-thick AlGaAs and a 400-nm-thick GaAs separate confinement heterostructure (SCH) layer. Subsequently, a MQW active region consists of 2 pairs of 5-nm-thick InGaAs well layers and 15-nm-thick GaAs barrier layers was deposited. A 100-nm-thick AlGaAs and a 400-nm-thick GaAs SCH layer, a 1800-nm-thick p-AlGaAs cladding layer and a 300-nm-thick p+-GaAs contact layer were then grown on top of the MQW active region, as shown in Fig. 1. (a). For the slightly-doped TJ LD, we deposited the TJ structure on top of the conventional LD, as shown in Fig. 1. (b). It should be noted that the growth condition for the TJ LD structure is identical to that for the conventional LD. On the other hand, the TJ structure consists of a 10-nm-thick n++-GaAs layer, a 8-nm-thick undoped GaAs layer, as shown in Fig. 1. (c). After the growth, standard Fabry-Perot laser procedures were then used to fabricate a ridge waveguide with a width of 4 μm and a depth of 2.2 μm. A silicon dioxide insulating layer was grown by Plasma-Enhanced Chemical Vapor Deposition (PECVD), Then Ti/Au metal p-contacts and Au/Ge/Ni n-contacts were deposited. Finally, the wafers were cleaved into L 300 μm × W 200 μm individual chips with both facets left uncoated (as-cleaved).

Compared to the conventional LD, the TJ LD has a large threshold current and low slope efficiency. The current threshold of the TJ LD is 36 mA, which is 3.6 times higher than conventional LD. With an output power of 20 mW, it was found that injection currents for the conventional LD and TJ LD were 82 mA and 172 mA, respectively, as shown in Fig. 2. The output power ($P_0$) above the threshold current ($I_{th}$) can be expressed as [16]:

$$P_0 = \eta_d \frac{h\nu}{q}(I - I_{th}) \quad (1)$$

$$\frac{1}{\eta_d} = \frac{\langle \alpha_i \rangle}{\eta_i \ln\frac{1}{R}} L + \frac{1}{\eta_i} \quad (2)$$

Where $\eta_d$ is the differential quantum efficiency, also known as the slope efficiency of the laser, $h\nu$ is the energy per photon, $q$ is the electronic charge, $I_{th}$ is the threshold current, $\eta_i$ is the internal quantum efficiency, $\langle \alpha_i \rangle$ is the average internal loss, $L$ is the cavity length of the laser, $R$ the facet reflectivity, and $R \cong 0.33$.

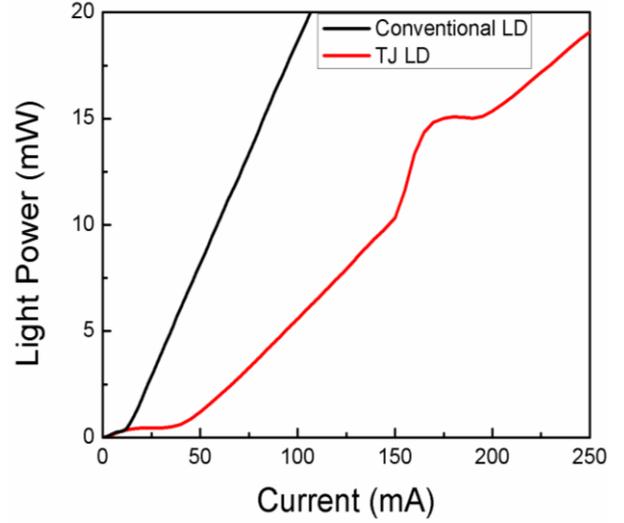

Fig. 2. P-I characteristics of the conventional LD (black line), and the TJ LD (red line).

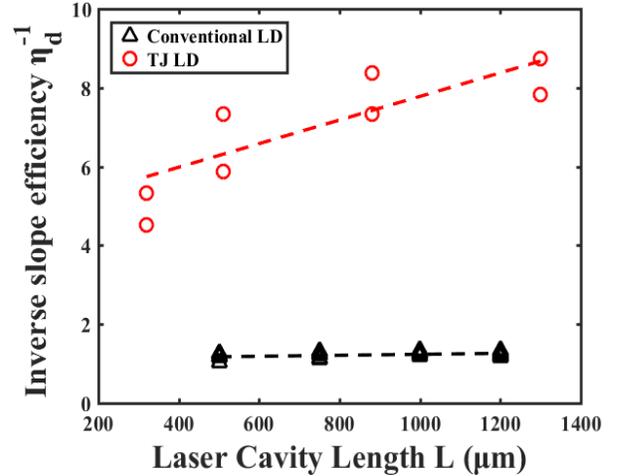

Fig. 3 Inverse slope efficiency $1/\eta_d$ versus laser length L as measured. Linear regression results are given as lines together with the parameters of Eq. (2).

Equation (2) gives a linear dependence and it is widely used to determine the average internal loss and internal quantum efficiency, by measuring the differential quantum efficiencies for different cavity lengths [16]. The results of the conventional LD and the TJ LD are shown in Fig. 3. Conventional LD internal quantum efficiency is 87.3% and the average internal loss is 0.85 cm[-1]. TJ LD internal quantum efficiency is 21% and the average internal loss is 6.9 cm[-1]. The low internal quantum efficiency and the high average internal loss lead to the low output power and the high current threshold. This is because the slightly-doped TJ increased the barrier width and free carrier absorption. As a result, the electron tunneling probability reduces, the effective resistivity of the TJ and the internal loss increase. High-resistive layers can be the reason for a strong lateral current spreading in the laser and, as a consequence, lower differential quantum efficiency and higher threshold currents.

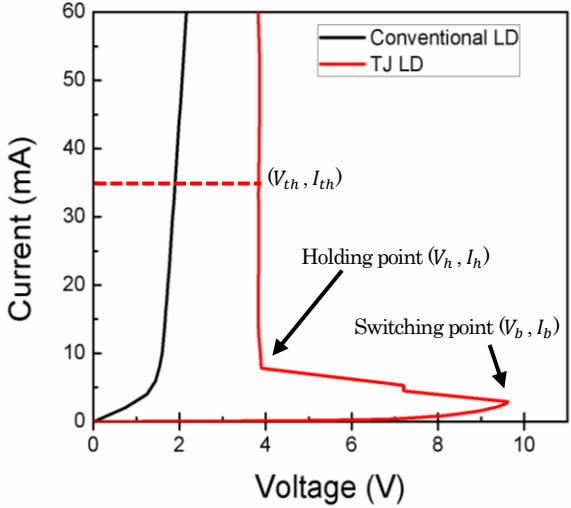

(a)

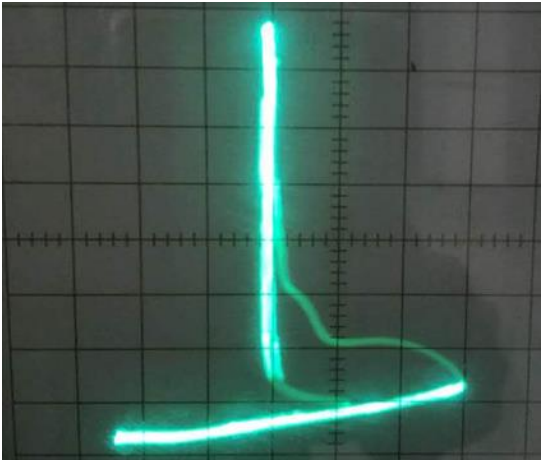

(b)

Fig. 4. (a) The I-V characteristics of the conventional LD (black line), the TJ LD (red line), and red dotted horizontal line represents threshold current of the TJ LD. (b) The I-V characteristic of the TJ LD measured by a homemade XJ4810 oscilloscope (curve tracer).

Current-Voltage (I-V) characteristics measured from these devices were plotted in Fig. 4. (a). It was found that the conventional LD has a normal p-n diode characteristic with the threshold voltage of 1.2 V. But for the TJ LD in the forward bias, it shows a nonlinear S-shape I-V characteristic, and has three distinct states: (1) high-impedance forward blocking region (off-state), (2) negative-resistance region, and (3) low-impedance and high-conductance forward-conducting region (on-state). In general, the PN junction grown by MOCVD is regarded as being approximately abrupt The reverse-biased avalanche breakdown voltage $V_{b0}$ of GaAs abrupt junction is given by the following formula [17]:

$$V_{b0} = 60(\frac{E_g}{1.1})^{3/2}(\frac{N_b}{10^{16}})^{-3/4} \qquad (3)$$

Where $E_g$ is the bandgap of GaAs material, and $N_b$ is the ionised impurity concentration in cm-3 in the lightly doped region. When the growth temperature of undoped GaAs is 780 ℃, the background doping impurities are mainly C, with a value of about $3 \times 10^{17}$ cm-3. $V_{b0}$ is 6.9 V calculated based on Eq. (3). The measured block voltage ($V_b$) and current ($I_b$) are 9.6 V and 3 mA, respectively. The holding voltage ($V_h$) and current ($I_h$) are 3.9 V and 7.8 mA, respectively. The lasing device with an S-shaped I-V characteristic can dedicate to the optical switching application [18]. An electric hysteresis loop of the TJ LD structure measured by a homemade XJ4810 oscilloscope (curve tracer), as shown in Fig. 4. (b). When the bias voltage increases from 0 to $V_b$, the on-state will begin after voltage touched $V_b$. However, when the bias voltage decreases, the on-state will begin as soon as over voltage $V_h$. In the on-state, the optical thyristor can emit light as a laser as long as the current is greater than the threshold. Consequently, this device has the potential to provide output power bisability. The turn-on voltage of the TJ LD is $V_{th}$, which almost close to $V_h$. When the voltage increases from 0 to $V_b$, the QW laser is in OFF state. However, when the voltage decreases from V (V>$V_b$) to $V_{th}$, the QW laser is in ON state. It is worth mentioning that the QW laser can turn on immediately [9][19].

Fig. 5. (a) and (b) show electroluminescence (EL) spectra measured from the conventional LD and the slightly-doped TJ LD, respectively. It can be seen that EL peaks occurred indeed at around 1060 nm for both LDs, as designed. It was also found that the peak wavelength of the conventional LD shifts from 1055 nm to 1058 nm with increasing injection current, which hints that wavelength tenability can be easily achieved not only through a temperature change, but also by adjusting the operation current. But for the slightly-doped TJ LD, it shows the broadband emission characteristic, as the injection level is increased to 4.5 × $I_{th}$, the spectrum broadens toward a longer wavelength and gives a 15 nm broad spectrum. This is because the lasing dynamics reflects the current dynamics formed as a result of complex nonlinear couplings within the laser-thyristor heterostructure [20]. The spectrum can be further broad with the increased injection current [21].

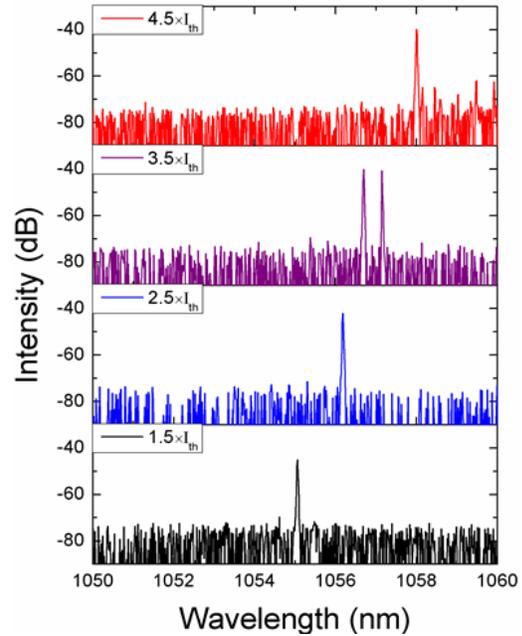

(a)

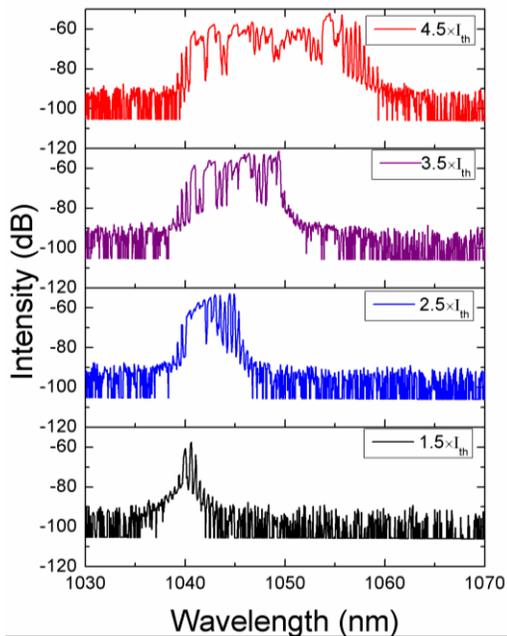
(b)

Fig. 5. EL spectra measured under different current from (a) the conventional LD, (b) the TJ LD.

In summary, We experimentally investigate and analyze the electrical and optical characteristics of GaAs-based conventional quantum well laser diodes and the quantum well laser diodes with slightly-doped tunnel junction. It was found that TJ LD show a nonlinear S-shape I-V characteristic.It was also found that the internal quantum efficiency measured by 21% and 87.3% for the TJ LD and the conventional LD, respectively. Furthermore, compared with the conventional LD, it was found that we could achieve 15 nm broadband spectrum from the TJ LD due to lasing dynamics reflects the current dynamics. The results may also lead to the realization of more applications, such as mode locking with ultra-short pulse generation, biomedical imaging, broad wavelength tenability, and multiwavelengths generation.

This work was supported by the National Key Research and Development Program of China under Grant No. 2017YFB0405301, and the National Natural Science Foundation of China under Grant No. 61604144, Grant No. 61504137.